\documentclass[draftcls,onecolumn,12pt]{IEEEtran}
\usepackage{epsfig}
\usepackage{amsmath,amssymb,amsfonts}
\usepackage{graphicx}
\usepackage{subfigure}

\newcommand{\beq}{\begin{equation}}
\newcommand{\eeq}{\end{equation}}
\newcommand{\beqn}{\begin{eqnarray}}
\newcommand{\eeqn}{\end{eqnarray}}

\newcommand{\vol}{\mathop{\rm vol}\nolimits}
\newcommand{\diag}{\mathop{\rm diag}\nolimits}

\newcommand{\mbf}[1]{\mbox{\boldmath $#1$}}
\newcommand{\R}{{\mathbb R}}

\newcommand{\Z}{{\mathbb Z}}
\newcommand{\EE}{{\mathbb E}}
\newcommand{\eqdef}{\stackrel{\Delta}{=}}

\newcommand{\Ac}{{\cal A}}
\newcommand{\Bc}{{\cal B}}
\newcommand{\Ec}{{\cal E}}
\newcommand{\Nc}{{\cal N}}

\newcommand{\Sc}{{\cal S}}
\newcommand{\Vc}{{\cal V}}
\newcommand{\Am}{{\bf A}}

\newcommand{\Hm}{{\bf H}}
\newcommand{\Qm}{{\bf Q}}
\newcommand{\Rm}{{\bf R}}
\newcommand{\Mm}{{\bf M}}

\newcommand{\Id}{{\bf I}}

\newcommand{\hv}{{\bf h}}
\newcommand{\uv}{{\bf u}}

\newcommand{\wv}{{\bf w}}
\newcommand{\xv}{{\bf x}}
\newcommand{\yv}{{\bf y}}
\newcommand{\zv}{{\bf z}}
\newcommand{\zerov}{{\bf 0}}
\newcommand{\gammav}{{\mbf \gamma}}
\newcommand{\alphav}{{\mbf \alpha}}
\newtheorem{theorem}{Theorem}

\newcommand{\openone}{\leavevmode\hbox{\small1\normalsize\kern-.33em1}}

\title{Sphere Lower Bound for Rotated Lattice Constellations in Fading Channels}

\author{Albert Guill\'en i F\`abregas
\thanks{A. Guill\'en i F\`abregas was with the Institute for Telecommunications Research, University of South Australia, Mawson Lakes SA 5095, Australia. He is now with the Department of Engineering, University of Cambridge, Cambridge CB2 1PZ, UK, e-mail: {\tt guillen@ieee.org}.} 
and Emanuele Viterbo
\thanks{E. Viterbo was with the Dipartimento di Elettronica, Politecnico di Torino, 10129 Torino, Italy. He is now with the Dipartimento di Elettronica Informatica e Sistemistica, Universit\`a della Calabria, via P. Bucci, 87036 Rende, Italy, e-mail: {\tt viterbo@deis.unical.it}.}
\thanks{The work by A. Guill\'en i F\`abregas has been supported by the Australian Research Council (ARC) Communications Research Network (ACoRN) under ARC grant RN0459498, by ARC Grants DP0344856 and DP0558861 and by the University of South Australia Australian Competitive Grant Development Scheme. The work by E. Viterbo has been supported by the International Visiting Researcher Program of the University of South Australia and the by the STREP project No. IST-026905 (MASCOT) within the sixth framework programme of the European Commission.}
\thanks{This work has been presented in part at the 2006 IEEE International Symposium on Information Theory, Seattle, July 2006.}
}

\begin{document}
\maketitle

\begin{abstract}
We study the error probability performance of rotated lattice
constellations in frequency-flat Nakagami-$m$ block-fading channels. In
particular, we use the sphere lower bound on the underlying infinite
lattice as a performance benchmark. We show that the sphere lower
bound has full diversity. We observe that optimally rotated lattices
with largest known minimum product distance perform very close to
the lower bound, while the ensemble of random rotations is shown to
lack diversity and perform far from it. 
\end{abstract}
\newpage

%%%%%%%%%%%%%%%%%%%%%%%%%%%%%%%%%%%%%%%%%%%%%%%%%%%%%%%%%%%%%%%%%%%%%%%%%%%%%%
\section{Introduction}
\label{s:introduction}

In this letter, we study the family of full rate multidimensional
signal constellations carved from lattices in frequency-flat Nakagami-$m$ fading
channels with $N$ degrees of freedom. In particular, we consider the
{\em uncoded} case, i.e., no time redundancy is added to the
transmitted signal. Current best constellations are designed to
achieve full diversity and maximize the minimum product distance
\cite{BOV04,FNT04,rotations_golden_web}. To date,
there exists no benchmark to compare the performance of rotated
lattice constellations. Recent work (\cite{viterbo_belfiore_isit05})
gives an approximation to the error probability of multidimensional
constellations in fading channels, which
is not tight and does not always have full diversity.

In this letter, we use the sphere lower bound\footnote{Literature
commonly refers to such bound as sphere-packing bound. In order to
avoid possible confusion with lattice terminology, we will refer to
it as sphere lower bound, since its computation is not based on the
packing radius of the lattice \cite{conway_sloane}.} (SLB), as a
benchmark for the performance of such uncoded lattice constellations. The SLB
dates back to Shannon's work \cite{shannon_bound}, and gives a lower
bound to the error probability of spherical codes with a given
length in the additive white Gaussian noise (AWGN) channel. The
application of the SLB to infinite lattices and lattice codes was
studied in \cite{Viterbo95,Tarokh99} for the AWGN channel. This SLB
yields a lower bound to the error probability of infinite lattices
regardless of the lattice structure. An approximated SLB was derived
in \cite{vialle_boutros_allerton99} for spherical codes over the
Rayleigh fading channel. Fozunbal {\em et al.} \cite{McLaughlin03}
extended the SLB to coded communication over the multiple-antenna
block-fading channel. A remarkable result of \cite{McLaughlin03} is
that, for a fixed number of antennas and blocks, as the code length
grows, the SLB converges to the outage probability of the channel
with Gaussian inputs \cite{telatar}. Unfortunately, the outage
probability \cite{telatar,ozarow_shamai_wyner} and the SLB of
\cite{McLaughlin03} are very far from the actual error probability
of uncoded multidimensional constellations. Moreover, as the block
length increases, the performance of uncoded modulations degrades,
and therefore, the outage probability and the SLB of
\cite{McLaughlin03} are not very useful as performance benchmarks.

In this letter, we use the SLB of the infinite lattice as a benchmark
for comparing multidimensional constellations in the block-fading
channel. We first show that the
SLB of infinite lattice rotations for the block fading channel has
full diversity regardless of the block length. We illustrate that as
the block length increases, the SLB increases as well. We also show
that multidimensional constellations obtained by algebraic rotations
with largest minimum product distance obtained from pairwise error
probability criteria \cite{BOV04,FNT04,rotations_golden_web} perform
very close to the lower bound and that the ensemble of random
rotations does not achieve full diversity.

%%%%%%%%%%%%%%%%%%%%%%%%%%%%%%%%%%%%%%%%%%%%%%%%%%%%%%%%%%%%%%%%%%%%%%%%%%%%%%
\section{System model}
\label{s:system_model}
We consider a flat fading channel whose
discrete-time received signal vector is given by
\beq 
\yv_\ell = \Hm \xv_\ell + \zv_\ell, \;\;\; \ell=1,\cdots,L
\label{eq:model} 
\eeq
where $\yv_\ell\in\R^N$ is the $N$-dimensional real received signal
vector, $\xv_\ell\in\R^N$ is the $N$-dimensional real transmitted
signal vector, $\Hm = \diag(\hv)\in\R^{N\times N}$, with
$\hv=(h_1,\dotsc,h_N)\in\R^N$, is the flat fading diagonal matrix,
and $\zv\in\R^N$ is the noise vector whose samples are i.i.d. $\sim\Nc(0,\sigma^2)$ with pdf
\[
p(\zv) = (2\pi\sigma^2)^{-\frac{N}{2}}\exp\left(-\frac{\|\zv\|^2}{2\sigma^2}\right)
\]
We define the signal-to-noise ratio (SNR) as $\rho =1/\sigma^2$. A {\em frame} is composed of $L$, $N$-dimensional modulation symbols or of $NL$ channel uses. The case of complex signals
obtained from $2$ orthogonal real signals can be similarly modeled
by \eqref{eq:model} by replacing $L$ with $L'=2L$.

We assume that the fading matrix $\Hm$ is constant during one frame
and it changes independently from frame to frame. This corresponds
to the {\em block-fading channel} with $N$ blocks
\cite{ozarow_shamai_wyner}. We further assume perfect channel state
information (CSI) at the receiver, i.e., the receiver perfectly
knows the fading coefficients. Therefore, for a given fading
realization, the channel transition probabilities are given by
\[
p(\yv|\xv,\Hm) = (2\pi\sigma^2)^{-\frac{N}{2}}
\exp\left(-\frac{1}{2\sigma^2}\|\yv-\Hm\xv\|^2\right)
\]
Moreover, we assume that the real fading coefficients follow a Nakagami-$m$ distribution
\[
p_h(x) = \frac{2m^mx^{2m-1}}{\Gamma(m)} e^{-m x^2}
\]
where $m>0$ \footnote{The literature usually considers $m\geq0.5$ \cite{proakis}. However, the fading distribution is well defined and reliable communication is possible for any $0<m<0.5$.} and 
\[
\Gamma(x) \eqdef \int_0^{+\infty} t^{x-1} e^{-t} dt
\]
is the Gamma function \cite{abramowitz_stegun}.
We define the coefficients
$\gamma_n=h_n^2$ for $n=1,\dotsc,N$, which correspond to the fading
power gains with pdf 
\[
p_\gamma(x) = \frac{m^mx^{m-1}}{\Gamma(m)} e^{-m x}
\]
and cdf 
\[
P_\gamma(x) = 1-\overline\Gamma(mx,m),
\]
respectively, where 
\[
\overline\Gamma(a,x) \eqdef \frac{1}{\Gamma(a)}\int_x^{+\infty} t^{a-1} e^{-t} dt
\]
is the normalized incomplete Gamma function \cite{abramowitz_stegun}. By analyzing Nakagami-$m$ fading, we can recover the analysis for a large class of fading statistics, including
Rayleigh fading by setting $m = 1$ and Rician fading with parameter $K$ by setting $m =(K + 1)^2/(2K + 1)$ \cite{simon2000dco}.

\vspace{-5mm}
%%%%%%%%%%%%%%%%%%%%%%%%%%%%%%%%%%%%%%%%%%%%%%%%%%%%%%%%
\subsection{Multidimensional Lattice Constellations}
\label{s:lattices}
%%%%%%%%%%%%%%%%%%%%%%%%%%%%%%%%%%%%%%%%%%%%%%%%%%%%%%%%
We assume that the transmitted signal vectors $\xv$ belong to an
$N$-dimensional signal constellation $\Sc\subseteq\R^N$. We consider
signal constellations $\Sc$ that are generated as a finite subset of
points carved from the infinite lattice $\Lambda = \{ \Mm \uv : \uv \in \Z^N
\}$ with full rank generator matrix $\Mm\in\R^{N\times N}$
\cite{conway_sloane}. For normalization purposes we fix
$\det(\Mm)=1$. For a given channel realization, we define the {\em
faded lattice} seen by the receiver as the lattice $\Lambda' = \{ \Mm' \uv : \uv \in \Z^N
\}$, whose generator matrix is given by $\Mm'=\Hm\Mm$.

In order to simplify the {\em labeling} operation, constellations
are of the type $\Sc = \{ \Mm \uv + \xv_0: \uv \in\Z_M^N \}$, where
$\Z_M=\{0,1,\dotsc,M-1\}$ represents an integer PAM constellation, $\log_2(M)$ is the
number of bits per dimension and $\xv_0$ is an offset vector which
minimizes the average transmitted energy. Therefore, the rate of
such constellations is $R=\log_2 M$ bit/s/Hz. This is usually
referred to as full-rate uncoded transmission.

In order to avoid {\em shaping} loss it is convenient to use {\em
cubic} lattice constellations \cite{BOV04,FNT04}. This implies that $\Mm$ should be an orthogonal
matrix ($\Mm\Mm^T=\Id$) . Nevertheless, this is
not required in the calculation of the SLB.

\vspace{-5mm}
%%%%%%%%%%%%%%%%%%%%%%%%%%%%%%%%%%%%%%%%%%%%%%%%%%%%%%%%%%%%
\subsection{Maximum Likelihood Decoding Error Probability}
\label{s:ml}
%%%%%%%%%%%%%%%%%%%%%%%%%%%%%%%%%%%%%%%%%%%%%%%%%%%%%%%%%%%%
At a given $\ell$, a maximum likelihood (ML) decoder with perfect
CSI makes an error whenever 
\[
\| \yv_\ell-\Hm \wv\|^2 \leq \| \yv_\ell- \Hm \xv \|^2
\]
for some $\wv\in\Sc,\,\,\wv\neq \xv$. These inequalities define the
so called {\em decision region} around $\xv$ (see Figure \ref{fig:voronoicell}).
Under ML decoding, the {\em frame} error probability is then given
by 
\beq 
P_{\rm f}(\rho) = \EE[P_{\rm f}(\rho|\hv)] =
\EE\left[1-(1-P_{\rm s}(\rho|\hv))^L\right] 
\eeq 
where $P_{\rm f}(\rho|\hv)$ and $P_{\rm s}(\rho|\hv)$ are the frame and
$N$-dimensional symbol error probabilities for a given channel
realization and SNR $\rho$, where the average is taken over the fading distribution.
For a given constellation $\Sc$, we can write that
\[
P_{\rm s}(\rho|\hv) = \EE [P_{\rm s}(\rho|\xv,\hv)] =
\frac{1}{|\Sc|} \sum_{\xv\in\Sc} \int_{\yv\notin\Vc(\xv,\hv)}
\!\!p(\yv|\xv,\hv) d\yv
\]
where $\Vc(\xv,\hv)$ is the decision region or {\em Voronoi region}
for a given multidimensional lattice constellation point $\xv$ and
fading $\Hm$. Computing the
regions $\Vc(\xv,\hv)$ and the exact error probability is in general
a very hard problem. In this paper we use the SLB
\cite{shannon_bound} as a lower bound on $P_{\rm f}$.
We define the {\em diversity order} as the asymptotic (for large
SNR) slope of $P_{\rm f}$ in a log-log scale,
\beq
d = - \lim_{\rho\to\infty} \frac{\log P_{\rm f}(\rho)}{\log\rho}.
\eeq 
The diversity order is usually a function of the fading distribution and the signal constellation $\Sc$. In this paper, we show that the diversity order is the product of the signal constellation diversity and a parameter of the fading distribution. In particular, we say that a constellation $\Sc$ has {\em full diversity} if the ML decoder is able to decode correctly in presence of $N-1$ deep fades. 

\vspace{-5mm}
%%%%%%%%%%%%%%%%%%%%%%%%%%%%%%%%%%%%%%%%%%%%%%%%%%%%%%%%%%%%%%%%%%%%%%%%%%%%%%%%%%%%%%%
%%%%%%%%%%%%%%%%%%%%%%%%%%%%%%%%%%%%%%%%%%%%%%%%%%%%%%%%%%%%%%%%%%%%%%%%%%%%%%%%%%%%%%%
\section{Sphere Lower Bound of a Faded Lattice}
\label{s:slb}
%%%%%%%%%%%%%%%%%%%%%%%%%%%%%%%%%%%%%%%%%%%%%%%%%%%%%%%%%%%%%%%%%%%%%%%%%%%%%%%%%%%%%%%%
In this Section, we recall the basics of the SLB for infinite
lattices $\Sc=\Lambda$ \cite{Viterbo95,Tarokh99} and we apply it to
bound $P_{\rm f}(\rho)$. The first simplification stems from the geometrical uniformity of
lattices, which implies that \cite{Viterbo95,Tarokh99} 
\[
\Vc(\xv,\hv) = \Vc(\wv,\hv),~~  \forall \xv,\wv\in\Lambda, \xv\neq\wv
\]
namely, for a given fading realization, the Voronoi regions of all
lattice points are equal. Let $\Vc_\Lambda(\hv)$ denote such Voronoi
region of the faded lattice.  Therefore, and without loss of
generality, we safely assume the transmission of the all-zero
codeword, i.e., $\xv_\ell= \zerov ~,~ \ell=1,\dotsc,L$.  Then, the
error probability is given by \cite{conway_sloane} 
\beq 
P_{\rm f}(\rho) =1 - \EE\left[\left( 1 - \int_{\zv\notin\Vc_\Lambda(\hv)}p(\zv) d\zv\right)^L\right]. 
\eeq
Due to the circular symmetry of the Gaussian noise, replacing
$\Vc_\Lambda(\hv)$ by an $N$-dimensional sphere $\Bc(\hv)$ of the same
volume and radius $R(\hv)$ \cite{shannon_bound}, yields the corresponding SLB
on the lattice performance \cite{Viterbo95,Tarokh99} 
\beq 
P_{\rm f}(\rho) \geq P_{\rm
slb}(\rho) \eqdef 1 - \EE\left[\left( 1 - \int_{\zv\notin\Bc(\hv)}
p(\zv) d\zv\right)^L\right] 
\label{eq:slb} 
\eeq
Since the volume of $\Bc(\hv)$ is \cite{conway_sloane}
\[
\vol(\Bc(\hv)) = \frac{\pi^{\frac{N}{2}}R(\hv)^N}{\Gamma(\frac{N}{2}+1)},
\] 
equating it
to the fundamental volume of the lattice
(volume of the Voronoi region) given by 
\[
\vol(\Vc_\Lambda(\hv)) = \det (\Hm\Mm) = \prod_{n=1}^N h_n
\]
yields the sphere radius
\beq 
R(\hv)^2 = \frac{1}{\pi}\Gamma\left(\frac{N}{2}+1\right)^{\frac{2}{N}}\left(\prod_{n=1}^N\gamma_n\right)^{1/N}. 
\label{eq:radius}
\eeq
The probability that the noise brings the received point outside the
sphere in \eqref{eq:slb} is simply expressed as
\cite{shannon_bound,Viterbo95,Tarokh99}
\beq 
P_{\rm slb}(\rho) = 1 - \EE\left[\left( 1 -\overline{\Gamma}\left(\frac{N}{2},\frac{R(\hv)^2}{2}\rho\right)\right)^L\right]. 
\label{eq:slb_gamma}
\eeq
We are now ready for the following result, whose proof is given in the Appendix.
%%%%%%%%%%%%%%%%%%%%%%%%%%%%%%%%%%%%%%%%
\begin{theorem}
\label{th:slb} In a Nakagami-$m$ block-fading channel with $N$ fading blocks, the
SLB on the error probability given in
(\ref{eq:slb_gamma}) has diversity order $d=mN$ for any $L\geq 1$, i.e., full diversity.
% for any $L\geq 1$, i.e.,
%\[
%d = -\lim_{\rho \rightarrow \infty} \frac{\log P_{\rm slb}(\rho)
%}{\log \rho} = mN.
%\]
\end{theorem}
%%%%%%%%%%%%%%%%%%%%%%%%%%%%%%%%%%%%%%%%

The previous theorem asserts that the best lattice in a channel with
$N$ fading blocks cannot have diversity larger than $mN$, showing that the overall diversity order is the product of the channel diversity $m$ and the maximal signal constellation diversity $N$.
This result
is non-trivial, and very important for constellation design.
Pairwise error probability analysis yields that full diversity
lattices can achieve full diversity
\cite{BOV04,FNT04,rotations_golden_web}, but no converse based on
the lattice structure has been proved so far for any $L$. Clearly,
if we construct our signal constellation $\Sc$ as a subset of points
of an $N$-dimensional lattice, $\Sc$ cannot have diversity larger than $m$ times the lattice dimension $N$.

In order to evaluate \eqref{eq:slb_gamma}, we need to perform a
multidimensional numerical integral over the joint distribution of the vector 
\[
\gammav=(\gamma_1,\dotsc,\gamma_N).
\] 
However, by carefully observing the expression of
$R(\hv)^2$ given in \eqref{eq:radius}, we can see that we only need to know the pdf of the
product of fading coefficients. It is not difficult to show that the characteristic function of the
random variable
\[
\zeta=\log\left(\prod_{n=1}^N \gamma_n \right)= \sum_{n=1}^N \log\gamma_n
\]
is given by (see Appendix B for details)
\beq 
G_\zeta(f) = \left(\frac{m^{j2\pi f-1}}{\Gamma(m)}\Gamma(m-j2\pi f)\right)^N.
\eeq
For $N>1$ a closed form inverse transform of this
function is not available, but we can nevertheless compute the pdf
$p_\zeta(z)$ numerically by using an inverse fast Fourier transform
(FFT). As an example, Figures \ref{fig:figPslbN} show the SLB for $L=1$ for various values of $N$ and $m$. As anticipated by Theorem
\ref{th:slb}, the curves get steeper as $m$ or $N$ increase. Moreover,
Figure \ref{fig:slb} shows the SLB for
$L=1,10,100,1000$ and various values of $N$ and $m$. For a given $N$ and $m$, all curves have the same
diversity. Observe that as $L$ increases the SLB increases, in contrast to what happens in the {\em coded} case, where as $L$ increases, the SLB converges to the outage probability of the channel, as demonstrated in \cite{McLaughlin03}. We note that the SNR $\rho =1/\sigma^2$ is relative to
the infinite lattice with vol$(\Lambda)=1$, since the average
transmitted energy cannot be defined.

%%%%%%%%%%%%%%%%%%%%%%%%%%%%%%%%%%%%%%%%%%%%%%%%%%%%%%%%%%%%%%%%%%%%%%%%%%%%%%
\section{Performance of rotated lattices}
\label{s:perf_lattices}

In this section, we give a number of examples that use the SLB as a
benchmark for comparing some lattices obtained by algebraic
rotations, as explained in section \ref{s:lattices}. In particular,
we will use the best known or optimal algebraically rotated $\Z^N$
lattices in terms of largest minimum product distance
\cite{BOV04,FNT04,frederique_thesis,rotations_golden_web}. As we
shall see, these rotations perform very close to the lower bound.
Furthermore, we will show that the ensemble of random rotations does
not have full diversity. This highlights the role of specific constructions that
guarantee full diversity and largest minimum product distance for approaching the SLB.

To illustrate this, Figures \ref{fig:pe_N2_m}, \ref{fig:peN2}, \ref{fig:peN4} and \ref{fig:peN8}, compare the frame error probability $P_{\rm f}(\rho)$ of optimal
rotations with largest minimum product distance (see
\cite{BOV04,FNT04,frederique_thesis} for more information on optimal constructions) obtained by simulation of the infinite
lattice using a Schnorr-Euchner decoder \cite{Agrell}
with the $P_{\rm slb}(\rho)$. The corresponding
rotation matrices are also available in \cite{rotations_golden_web}\footnote{Remark that the rotations in \cite{rotations_golden_web} are given in row format as in \cite{conway_sloane} and that here we use the column convention for lattice generator matrices.}. In particular, Figure \ref{fig:pe_N2_m} compares the performance of the cyclotomic rotation for $N=2$ and $L=1,100$ and $m=0.5,1,2$. Figures \ref{fig:peN4} and \ref{fig:peN8} show the SLB and the optimal rotations for $N=4,8$, namely the Kr\"uskemper and cyclotomic rotations respectively \cite{BOV04,FNT04,frederique_thesis}. As we observe, optimal rotations are very close to the SLB. As $N$ increases, algebraic rotations with largest minimum product
distance show some gap to $P_{\rm slb}(\rho)$. This is due to the
fact that for large $N$, the minimum product distance is not the
only relevant design parameter for optimizing the coding gain. Without any loss of generality in the presentation of our results, from now on, and unless otherwise specified, forthcoming examples will be shown for $m=1$.

Figures \ref{fig:averageHaar}, \ref{fig:peN4} and \ref{fig:peN8} also compare by simulation the performance of the aforementioned full-diversity
algebraic rotations with the average performance of the ensemble of
random rotations. To compute it, at every frame we generate a random matrix $\Am$
with zero mean and unit variance i.i.d. Gaussian entries. We then
perform a $\Am=\Qm\Rm$ decomposition and let $\Mm=\Qm$. This is
the simplest way of generating the ensemble of random rotations
(orthogonal matrices) with the Haar distribution \cite{stewart,tulino2004rmt}. As
we observe, algebraic rotations perform very close to $P_{\rm
slb}(\rho)$. On the other hand, the average error probability over
the ensemble of random rotations, lacks of the full diversity and
shows bad performance. To better understand this behavior, Figure \ref{fig:random30rot} shows the simulated performance of 30 random samples of the Haar ensemble for $N=1$ and $L=1$, compared to the SLB (thick solid line), performance of the cyclotomic rotation (circles) and the ensemble average (thick dashed line). We observe that almost all instances have full diversity (though with very different coding gains). However, the ensemble average performance is dominated by {\em bad} rotation matrices. In particular, a closer look to the two worse curves reveals that the corresponding rotation matrices are very close to the identity, achieving effectively no rotation nor diversity. Furthermore, we observe that as $N$ increases, the performance of random rotations improves, despite showing a different asymptotic slope. This is due to the fact that for large $N$, there is a lot of diversity in the channel and the error probability curves get very steep. This means that for large $N$, random rotations will perform well for low-to-medium SNR.

%%%%%%%%%%%%%%%%%%%%%%%%%%%%%%%%%%%%%%%%%%%%%%%%%%%%%%%%%%%%%%%%%%%%%%%%%%%%%%
\section{Performance of multidimensional signal sets}
\label{s:perf_sets}

%So far, we have seen that the performance of algebraic rotations is
%very close to the lower bound $P_{\rm slb}(\rho)$. 
Practical systems use finite signal alphabets and the
performance of the infinite rotated lattice should serve mainly as a
guideline. Unfortunately, we do not have a bound similar to $P_{\rm
slb}(\rho)$ for the finite case to take into account the boundary
effects. We conjecture that the best multidimensional signal set using $M$-PAM is the one that has generator matrix $\Mm$ such that $P_{\rm f}(\rho)$ is closest to $P_{\rm slb}(\rho)$ for large enough $\rho$.
As we shall see in the following example, as $M$ increases, the
performance of the multidimensional signal constellation approaches
that of the infinite rotated lattice, despite the boundary effects.
This is precisely the continuity argument used in \cite{Tarokh99} for lattice codes.
Indeed, Figures \ref{fig:peN2pam}, \ref{fig:peN4pam} and \ref{fig:peN8pam} show the
performance for $N=2,4,8$ and $L=1,100$ of the signal constellations
obtained from $M$-PAM with the optimal algebraic rotation. In the
comparison with the infinite lattice (circles) and $P_{\rm
slb}(\rho)$, we observe all curves are within $1.5$ dB.
Note that the SNR axis does not take into account the different
average energies of the finite constellations and that we assume
that the minimum distance of the $M$-PAM is $1$ for comparison to
the infinite lattice lower bound. In order to plot the performance
in terms of $\frac{E_b}{N_0}=\frac{E_b}{2}\rho$ it is enough to
shift the curves by 
\[
10\log_{10}\left(\frac{M^2-1}{24\log_2 M}\right) \rm dB
\]
%

%%%%%%%%%%%%%%%%%%%%%%%%%%%%%%%%%%%%%%%%%%%%%%%%%%%%%%%%%%%%%%%%%%%%%%%%%%%%%%
\section{Conclusions}
\label{s:conclusions} In this paper we have studied the performance
of multidimensional rotated lattice constellations. We have applied
the sphere lower bound for the infinite lattice to the block-fading
channel and proved that the bound has full diversity. We have shown
that optimally rotated algebraic lattices perform very close to the
bound, while the average over the ensemble of random rotations does
not. Furthermore, we have shown that finite constellations obtained
from the rotation of $\{M\!-\!\mbox{PAM}\}^N$ constellations perform
close to the bound as $M$ gets large. We have conjectured that
optimal multidimensional signal sets with  $M$-PAM constellation are
obtained from rotated lattices whose performance is closest to the
sphere lower bound.

%%%%%%%%%%%%%%%%%%%%%% APPENDIX %%%%%%%%%%%%%%%%%%%%%%%%%%%%%%%%%%%%%%%%%%%%%%

%%%%%%%%%%%%%%%%%%%%%%%%%%%%%%%%%%%%%%%%%%%%%%%%%%%%%%%%%%%%%%%%%%%%%%%%%%%%%%
\newpage
\appendices
\section*{Appendix A: Proof of Theorem \ref{th:slb}}
\label{appendix:proof_th_bfc}

The exponential equality $\doteq$ and inequalities $\dot\geq$ and $\dot\leq$ were introduced in \cite{Tse}. We write $f(z) \doteq z^d$ to indicate that $\lim_{z\rightarrow \infty} \frac{\log f(z)}{\log z} =d$. The exponential inequalities $\dot\geq$ and $\dot\leq$ are defined similarly. The function $\openone\{\Ec\}$ is the indicator function of the event $\Ec$, namely, $\openone\{\Ec\}=1$ when $\Ec$ is true, and zero otherwise. Following \cite{Tse}, we define the normalized
fading gains $\alpha_n\eqdef-\frac{\log \gamma_n}{\log\rho}$. 
It is not difficult to show that the joint pdf of the vector
$\alphav=(\alpha_1,\dotsc,\alpha_N)$ is given by,
\[
p(\alphav) = \left(\frac{m^m\log\rho}{\Gamma(m)}\right)^N e^{-m\sum_{n=1}^N \rho^{-\alpha_n}}
\rho^{-m\sum_{n=1}^N \alpha_n}.
\]
Using the same arguments as in \cite{Tse,albert_beppe_it,nguyen2007} we have that asymptotically for large $\rho$
\[
p(\alphav) \doteq \rho^{-m\sum_{n=1}^N \alpha_n}
\]
for $\alphav\in\R^N_+$, where $\R_+$ are the positive reals including zero.
We can express the SLB as,
\beq
P_{\rm slb}(\rho) = 1 - \int_{\R^N} \left[ 1 -
\overline{\Gamma}\left(\frac{N}{2},\beta(\alphav) \right)\right]^L
p(\alphav)d\alphav
\eeq
where
\beq
\beta(\alphav) \eqdef
\frac{1}{2\pi} \Gamma\left(\frac{N}{2}+1\right)^{\frac{2}{N}}
\rho^{1-\frac{1}{N}\sum_{n=1}^N \alpha_n}
\eeq 
is the second argument of the incomplete Gamma function in \eqref{eq:slb_gamma} as a function of $\alphav$.
Since $0\leq\left[ 1 - \overline{\Gamma}\left(\frac{N}{2},\beta(\alphav)
\right)\right]^L\leq1$ we can apply the dominated convergence theorem \cite{durrett_book}
and write
\begin{align}
\displaystyle\lim_{\rho\to\infty} \int_{\R^N} \left[ 1 -
\overline{\Gamma}\left(\frac{N}{2},\beta(\alphav) \right)\right]^L
p(\alphav)d\alphav \nonumber ~=~ \displaystyle\int_{\R^N} \lim_{\rho\to\infty} \left[ 1 -
\overline{\Gamma}\left(\frac{N}{2},\beta(\alphav) \right)\right]^L
p(\alphav)d\alphav.
\end{align}
Therefore, since
\begin{equation}
\lim_{\rho\to\infty}\beta(\alphav) =
\begin{cases}
0 & \text{if $\sum_{n=1}^N\alpha_n>N$}\\
\infty & \text{if $\sum_{n=1}^N\alpha_n<N$}
\end{cases}
\eeq we have that 
\beq \lim_{\rho\to\infty}
\overline{\Gamma}\left(\frac{N}{2},\beta(\alphav) \right) =
\begin{cases}
1 & \text{if $\sum_{n=1}^N\alpha_n>N$}\\
0 & \text{if $\sum_{n=1}^N\alpha_n<N,$}
\end{cases}
\eeq 
which implies that
\beq \lim_{\rho\to\infty} 1-\left[ 1- \overline{\Gamma}\left(\frac{N}{2},\beta(\alphav) \right) \right]^L =
\begin{cases}
1 & \text{if $\sum_{n=1}^N\alpha_n>N$}\\
0 & \text{if $\sum_{n=1}^N\alpha_n<N$}
\end{cases}
\eeq
and 
means that for any $L\geq 1$, the contribution to $P_{\rm slb}(\rho)$ from $\alphav$ such that $\sum_{n=1}^N\alpha_n<N$ is negligible for large $\rho$.
Also, since $p(\alphav) \dot=\rho^{-m\sum_{n=1}^N \alpha_n}$, we can write that, for
every $L\geq 1$, 
\beq 
P_{\rm slb}(\rho)= \int_{\alphav\in\R^N} p(\alphav) d\alphav ~ \doteq ~\int_{\alphav\in\Ac\cap\R^N_+} \rho^{-m\sum_{n=1}^N \alpha_n} d\alphav 
\eeq
where $\Ac = \left\{\alphav\in\R^N : \sum_{n=1}^N\alpha_n>N
\right\}$.
Therefore the diversity order of the SLB is
given by 
\beq 
d = -\lim_{\rho\to\infty} \frac{1}{\log\rho} \log
\int_{\alphav\in\Ac\cap\R^N_+} \exp\left(-m\log\rho \sum_{n=1}^N \alpha_n\right)
d\alphav 
\eeq 
We now apply Varadhan's lemma \cite{dembo_zeitouni}
and we obtain that 
\begin{align} 
d &= \inf_{\alpha\in\Ac\cap\R^N_+}\left\{m\sum_{n=1}^N\alpha_n\right\} = m\inf_{\alpha\in\Ac\cap\R^N_+}\left\{\sum_{n=1}^N\alpha_n\right\}= mN 
\end{align} 
which completes the proof.
%%%%%%%%%%%%%%%%%%%%%%%%%%%%%%%%%%%%%%%%%%%%%%%%%%%%%%%%%%%%%%%%%%%%%%%%%%%%%%

\newpage
%%%%%%%%%%%%%%%%%%%%%%%%%%
% APPENDIX II
\section*{Appendix B: Distribution of $\zeta$}
\label{appendix:distribution}
Consider the random variable $\gamma$ with pdf 
\[
p_\gamma(x) = \frac{m^mx^{m-1}}{\Gamma(m)} e^{-m x}
\]
for $x,m>0$, then the cdf of $\log \gamma$ can be expressed as
\begin{align}
\Pr(\log \gamma\leq x) &=P_\gamma(e^x)\\
& = 1-\overline\Gamma(m,me^x)
\end{align}
and the pdf of  $\log \gamma$ is given by
\[
p_{\log\gamma}(x) = \frac{m^m}{\Gamma(m)}e^{m(x-e^x)} ~~~~-\infty<x<\infty
\]
The corresponding characteristic function can be written as
\begin{align}
G(f) &=\EE[e^{-j2\pi f x}] \nonumber\\
& = \int_{-\infty}^{+\infty} e^{-j2\pi f x}\frac{m^m}{\Gamma(m)}e^{m(x-e^x)} \,dx
\end{align}
where using the change of variables $y=me^x$ yields
\begin{align}
G(f) & = \int_0^{+\infty}\frac{m^{j2\pi f -1}}{\Gamma(m)} y^{m-1-j2\pi f}e^{-y} dy\\
&=\frac{m^{j2\pi f-1}}{\Gamma(m)}\Gamma(m-j2\pi f).
\end{align}
Finally, the characteristic function of $\zeta$ is given by
\[
G_\zeta(f) =\left(\frac{m^{j2\pi f-1}}{\Gamma(m)}\Gamma(m-j2\pi f)\right)^N.
\]
Figure \ref{fig:figpzN} shows $p_\zeta(z)$ evaluated numerically. In particular, Figure \ref{fig:pdfm1} shows the $p_\zeta(z)$ for different values of $N$ and $m=1$ while Figure \ref{fig:pdfN8} shows $p_\zeta(z)$ for $N=8$ and different values of $m$.

%%%%%%%%%%%%%%%%%%%%%%%%%%%%%%%%%%%%%%%%%%%%%%%%%%%%%%%%%%%%%%%%%%%%%%%%%%%%%%
%\newpage

%%%%%%%%%%%%%%%%%%%%%%%%%%%%%%%%%%%%%%%%%%%%%%%%%%%%%%%%%%%%%%%%%%%%%%%%%%%%%%
%\textheight=9.40in
\newpage
 
\bibliographystyle{IEEE}

%%%%%%%%%%%%%%%%%%%%%%%%%%%%%%%%%%%%%%%%%%%%%%%%%%%%%%%%%%%%%%%%%%%%%%%%%%%%%%%
%%%%%%%%%%%%%%%%%%%%%%% FIGURES %%%%%%%%%%%%%%%%%%%%%%%%%%%%%%%%%%%%%%%%%%%%%%%
%%%%%%%%%%%%%%%%%%%%%%%%%%%%%%%%%%%%%%%%%%%%%%%%%%%%%%%%%%%%%%%%%%%%%%%%%%%%%%
\newpage

\begin{figure}
\includegraphics[width=1\columnwidth]{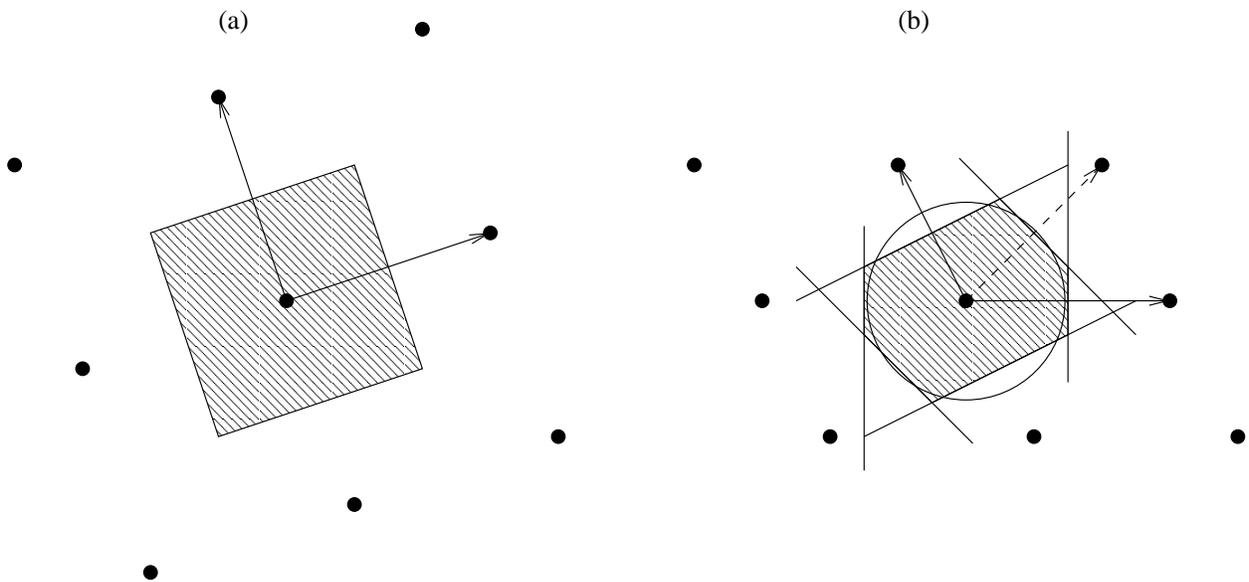}
\caption{The decision regions of the rotated $\Z^2$ lattice: (a)
before fading, (b) after fading $\hv=(1,0.5)$.}
\label{fig:voronoicell} 
\end{figure}

\begin{figure}[htbp]
  \centering
  \subfigure[$m=1$, $N=2,4,8,16,32,64$.]{\includegraphics[width=0.75\columnwidth]{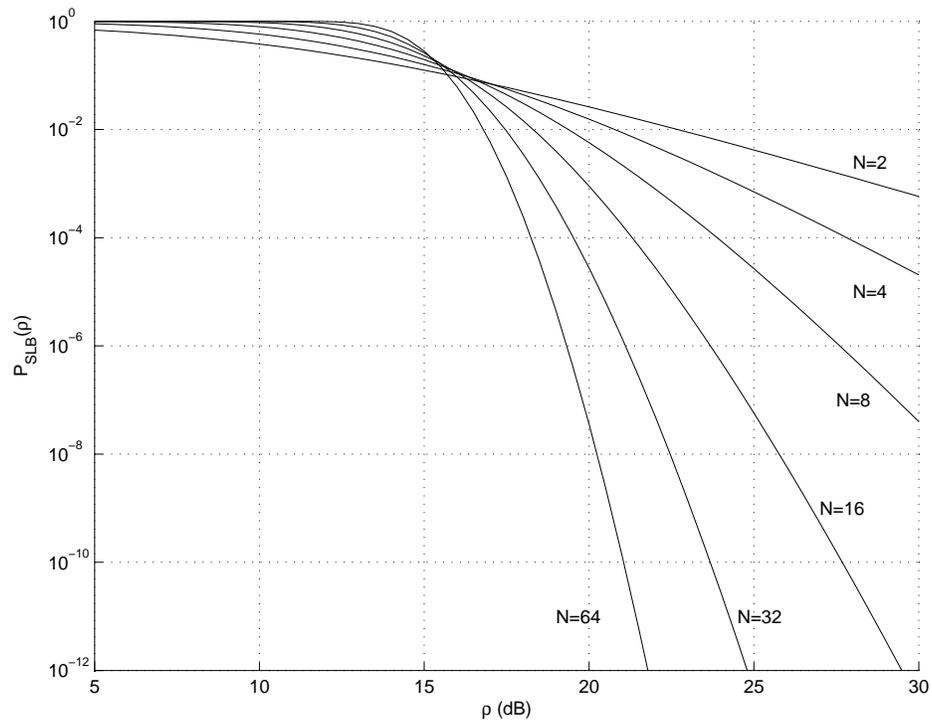}}
  \subfigure[$N=4$, $m=0.2,0.4,0.6,0.8,1,1.4,2$.]{\includegraphics[width=0.75\columnwidth]{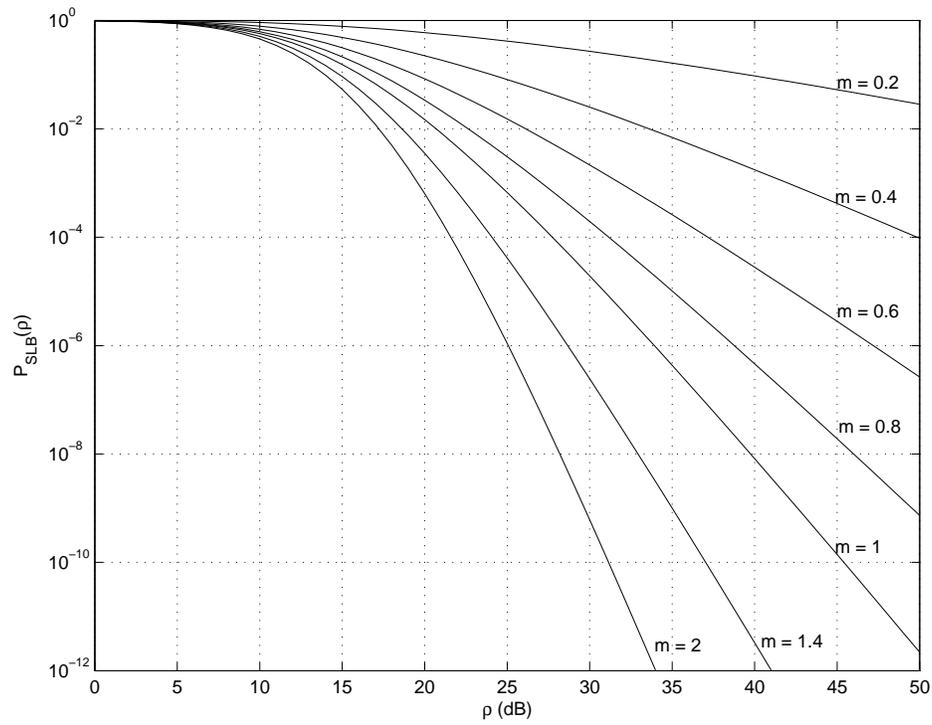}}
  \caption{Sphere lower bound $P_{\rm slb}(\rho)$ for various values of $N$ and $m$.}
  \label{fig:figPslbN}
\end{figure}

\begin{figure}[htbp]
  \centering
  \includegraphics[width=0.9\columnwidth]{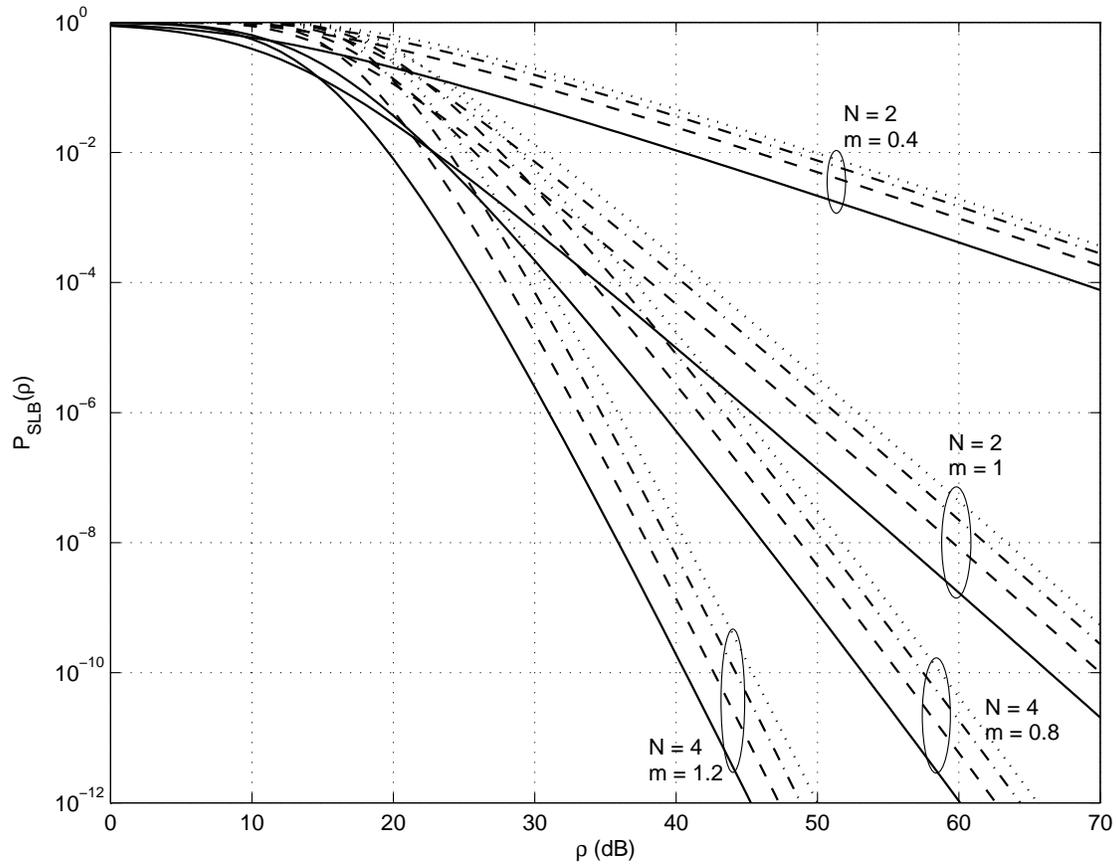}
  \caption{Sphere lower bound $P_{\rm slb}(\rho)$ for $L=1$ (solid), $L=10$ (dashed),
  $L=100$ (dashed-dotted) and $L=1000$ (dotted) and various values of $N$ and $m$.}
  \label{fig:slb}  
\end{figure}

\begin{figure}[htbp]
  \centering
  \includegraphics[width=0.9\columnwidth]{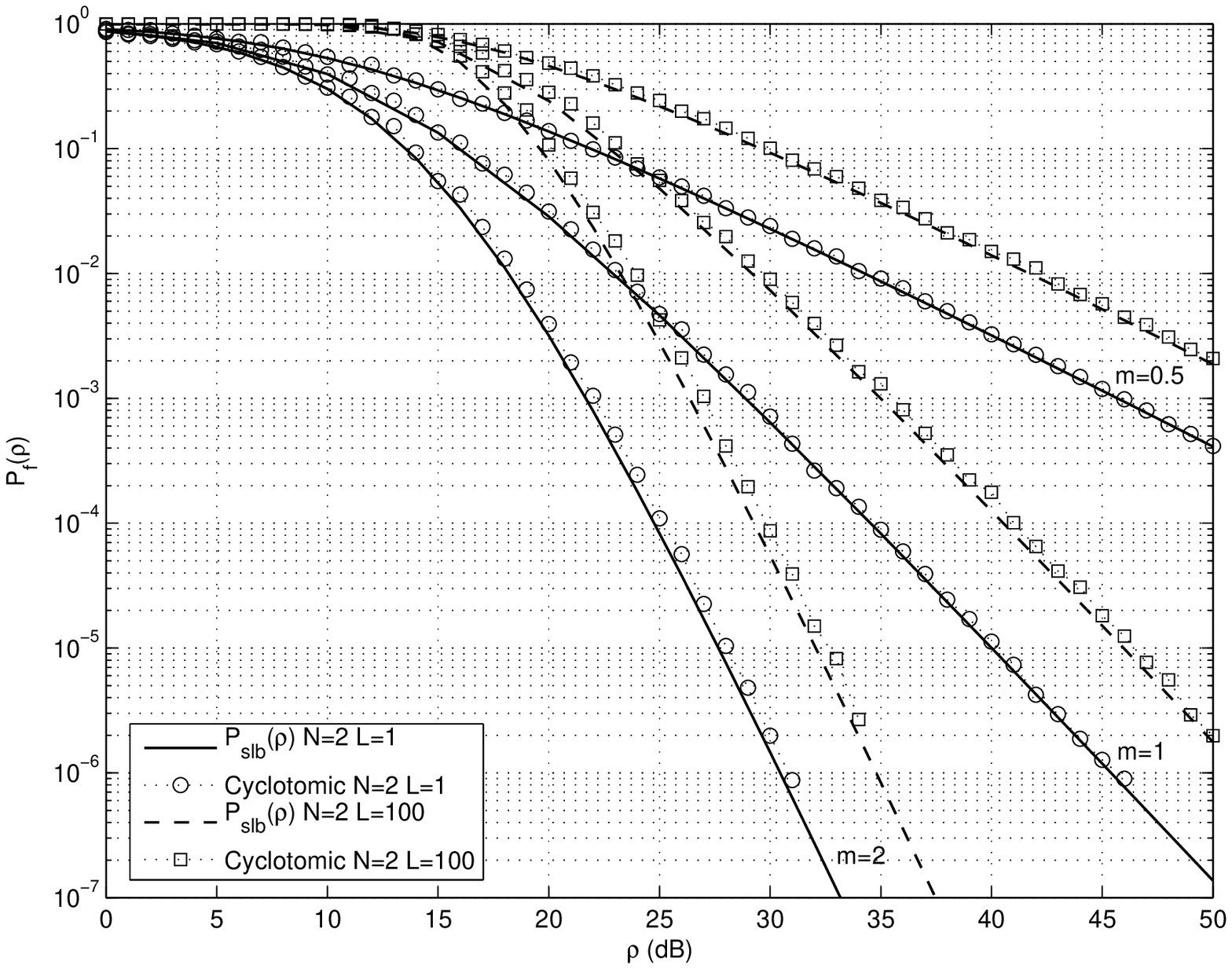}
  \caption{Frame error probability $P_{\rm f}(\rho)$ and sphere lower bound
  $P_{\rm slb}(\rho)$ for $N=2$, $m=0.5,1,2$ and $L=1,100$.}
  \label{fig:pe_N2_m}
\end{figure}

\begin{figure}[htbp]
  \centering
  \subfigure[\label{fig:averageHaar}Cyclotomic and average over Haar ensemble for $L=1,100$.]{\includegraphics[width=0.73\columnwidth]{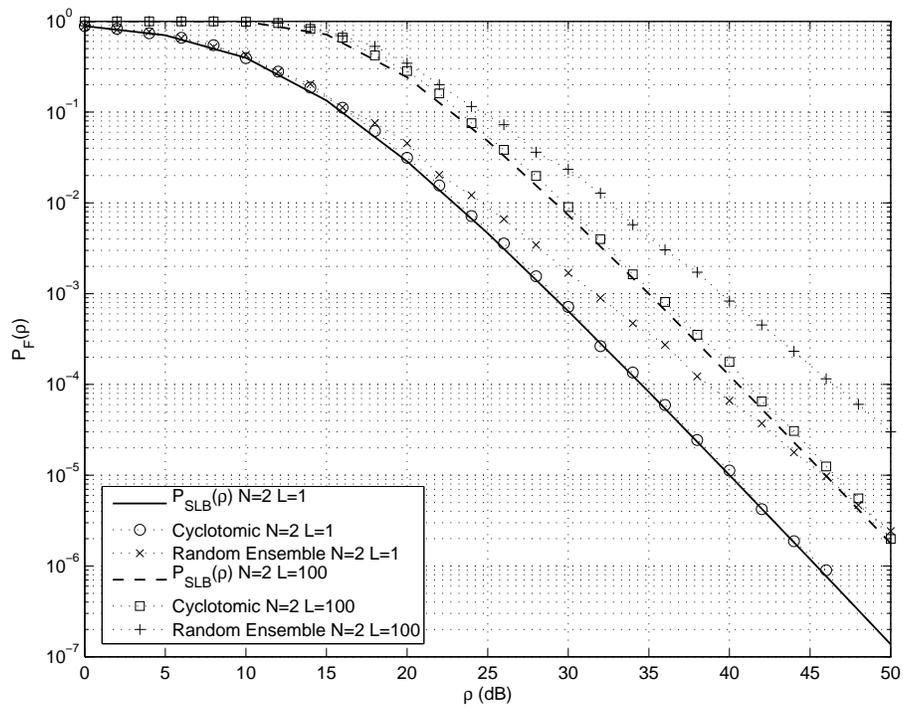}}
	\subfigure[\label{fig:random30rot}30 random samples from the Haar ensemble for $L=1$. The SLB (thick solid), average over random rotations (thick dashed) and Cyclotomic (circles) are shown for reference]{\includegraphics[width=0.73\columnwidth]{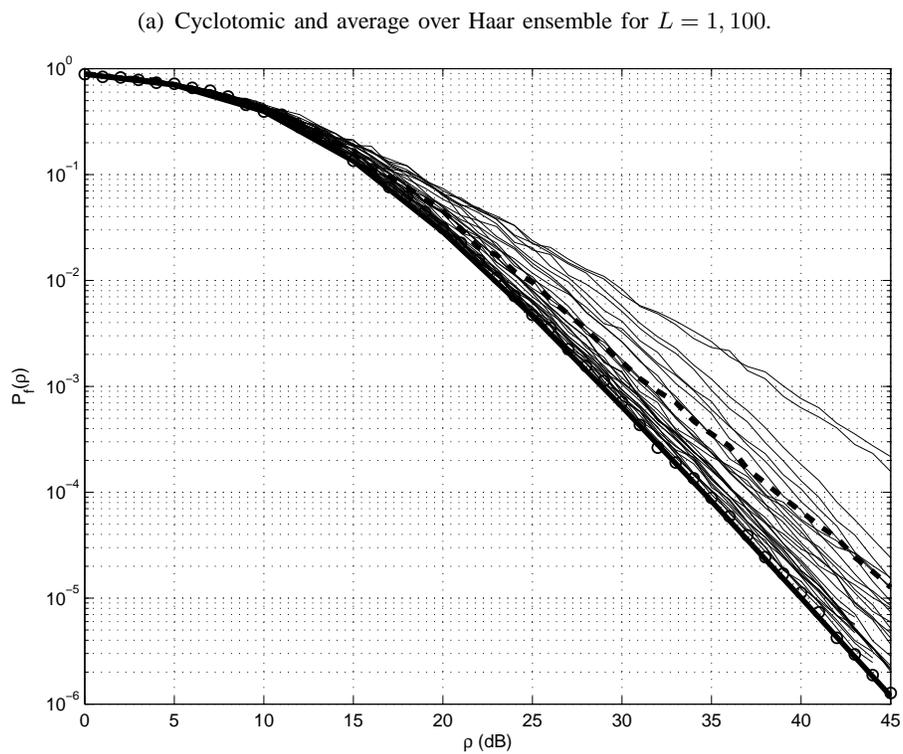}}
  \caption{Frame error probability $P_{\rm f}(\rho)$ and sphere lower bound
  $P_{\rm slb}(\rho)$ for $N=2$, $m=1$ and $L=1,100$.}
  \label{fig:peN2}
\end{figure}

\begin{figure}[htbp]
  \centering
  \includegraphics[width=0.9\columnwidth]{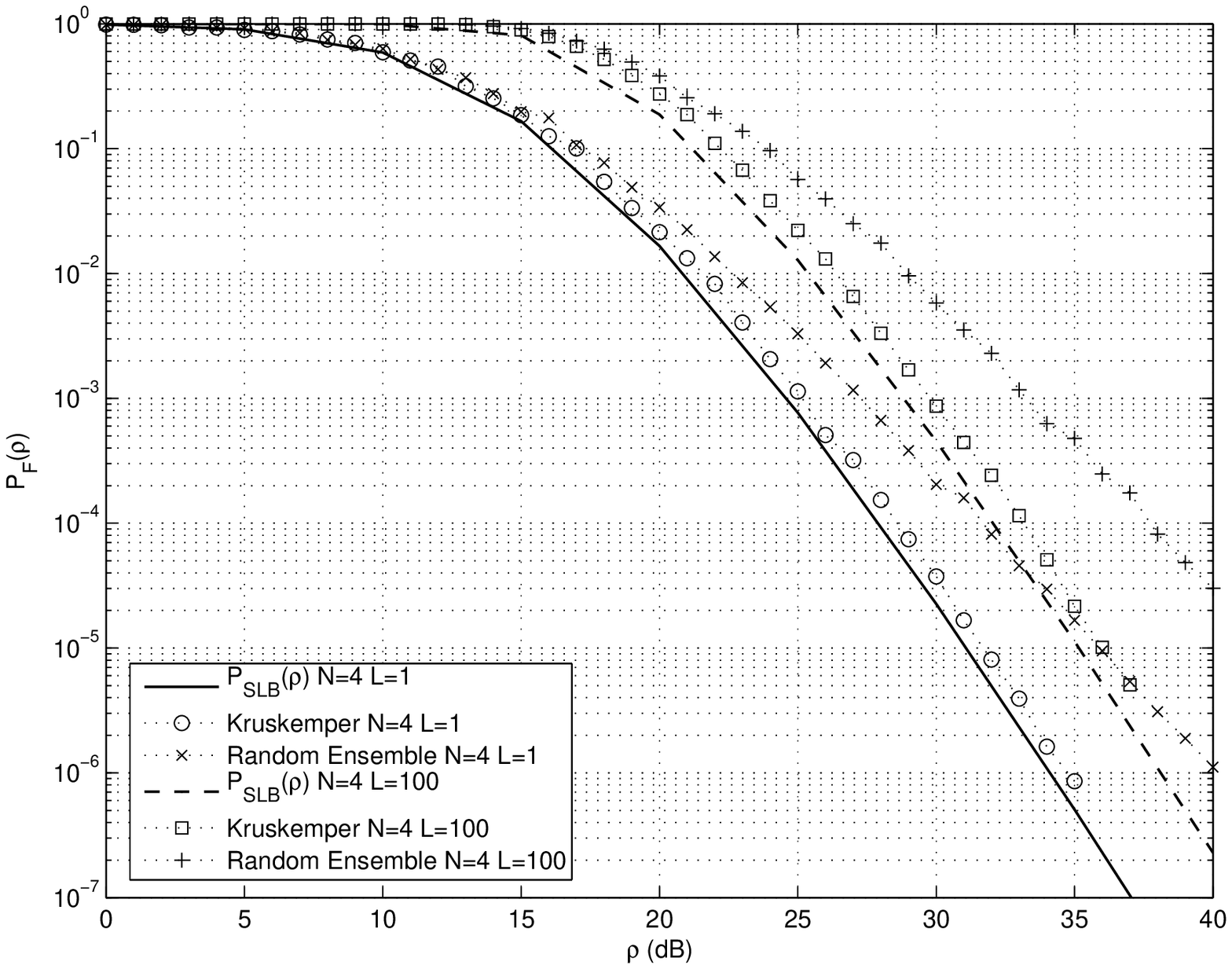}
  \caption{Frame error probability $P_{\rm f}(\rho)$ and sphere lower bound
  $P_{\rm slb}(\rho)$ for $m=1$, $N=4$, $L=1,100$ with Kr\"uskemper rotation.}
\label{fig:peN4} 
\end{figure}

\begin{figure}[htbp]
  \centering
  \includegraphics[width=0.9\columnwidth]{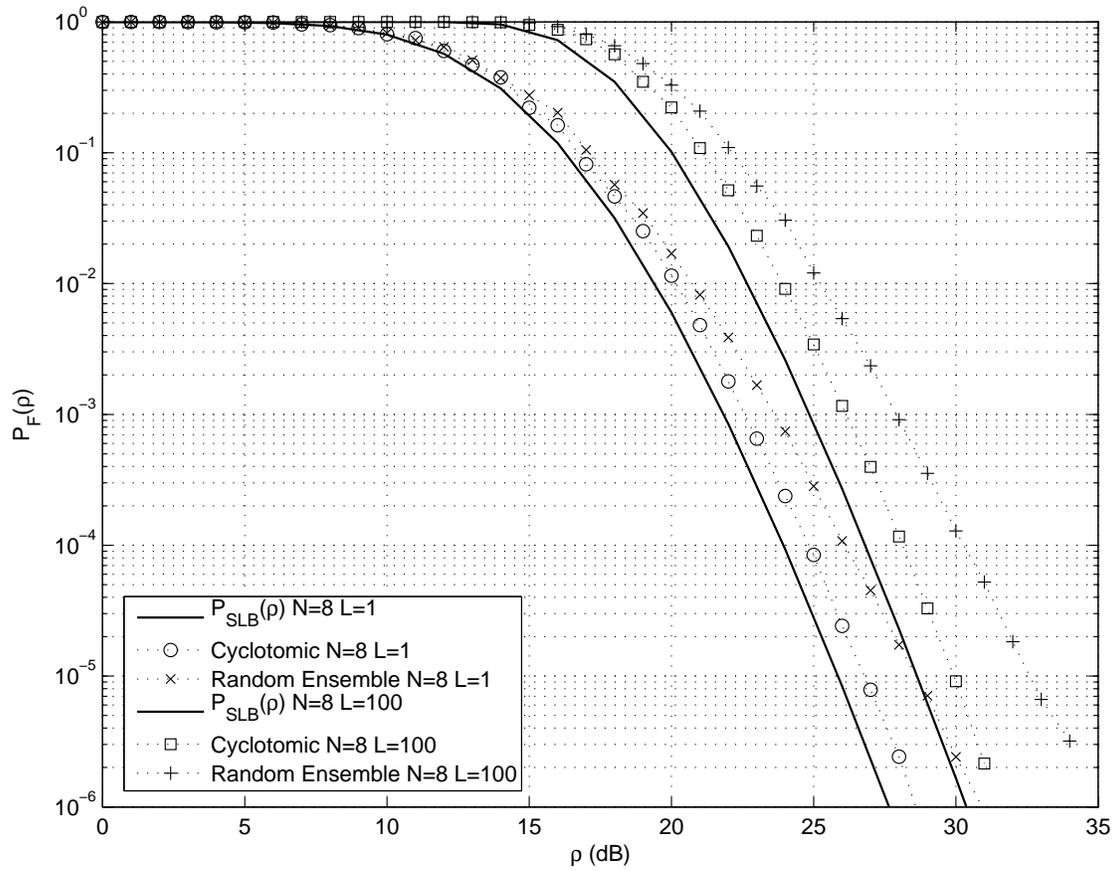}
  \caption{Frame error probability $P_{\rm f}(\rho)$ and sphere lower bound
  $P_{\rm slb}(\rho)$ for $m=1$, $N=8$, $L=1,100$ with cyclotomic rotation.}
\label{fig:peN8} 
\end{figure}

\begin{figure}[htbp]
  \centering
  \includegraphics[width=0.9\columnwidth]{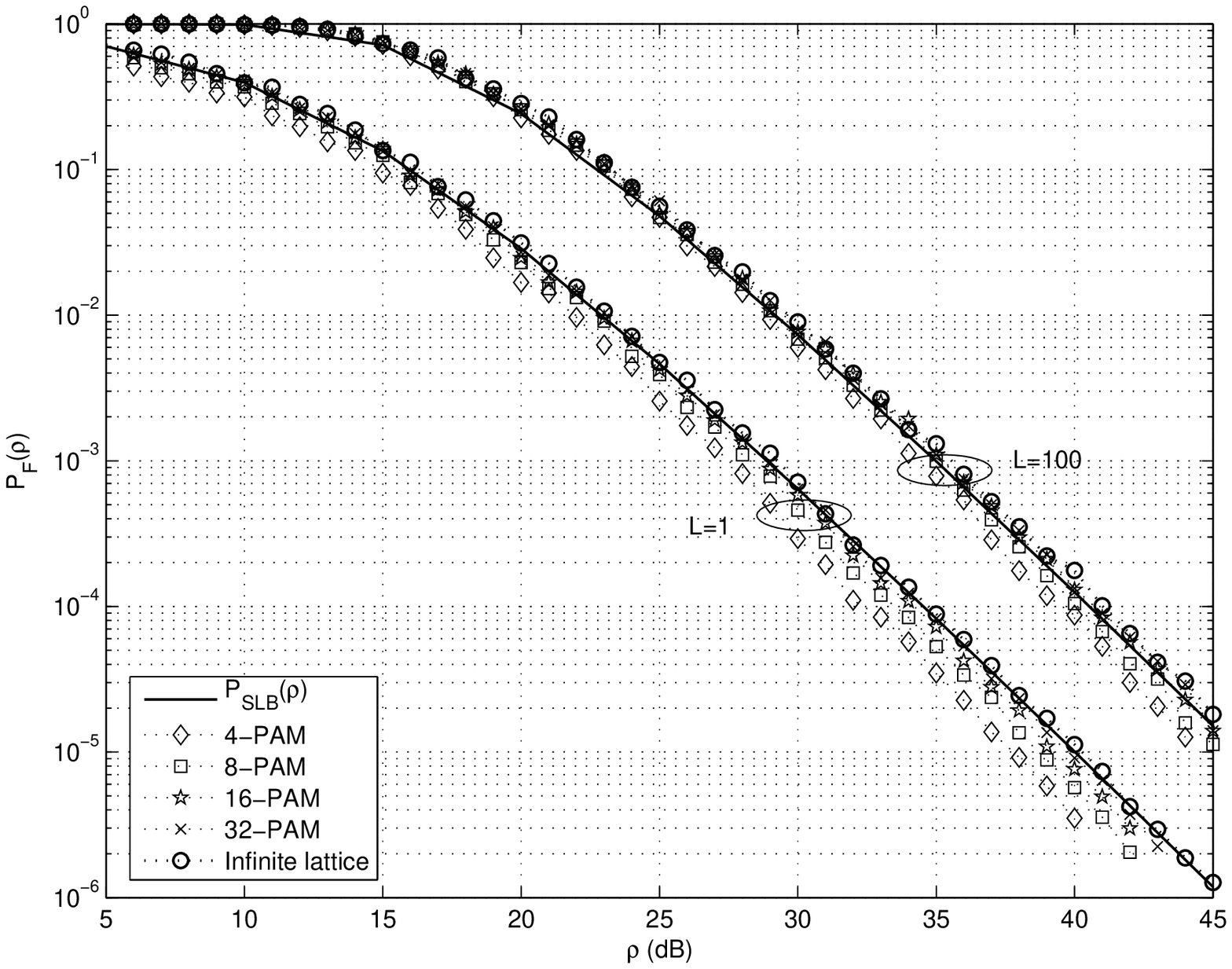}
  \caption{Frame error probability $P_{\rm f}(\rho)$ of the finite
  constellation generated with $4,8,16,32$-PAM, $P_{\rm f}(\rho)$
  with $N=2$ of the infinite lattice and sphere lower bound $P_{\rm slb}(\rho)$
  for $L=1,100$ with the cyclotomic rotation.}
  \label{fig:peN2pam}
\end{figure}

\begin{figure}[htbp]
  \centering
  \includegraphics[width=0.9\columnwidth]{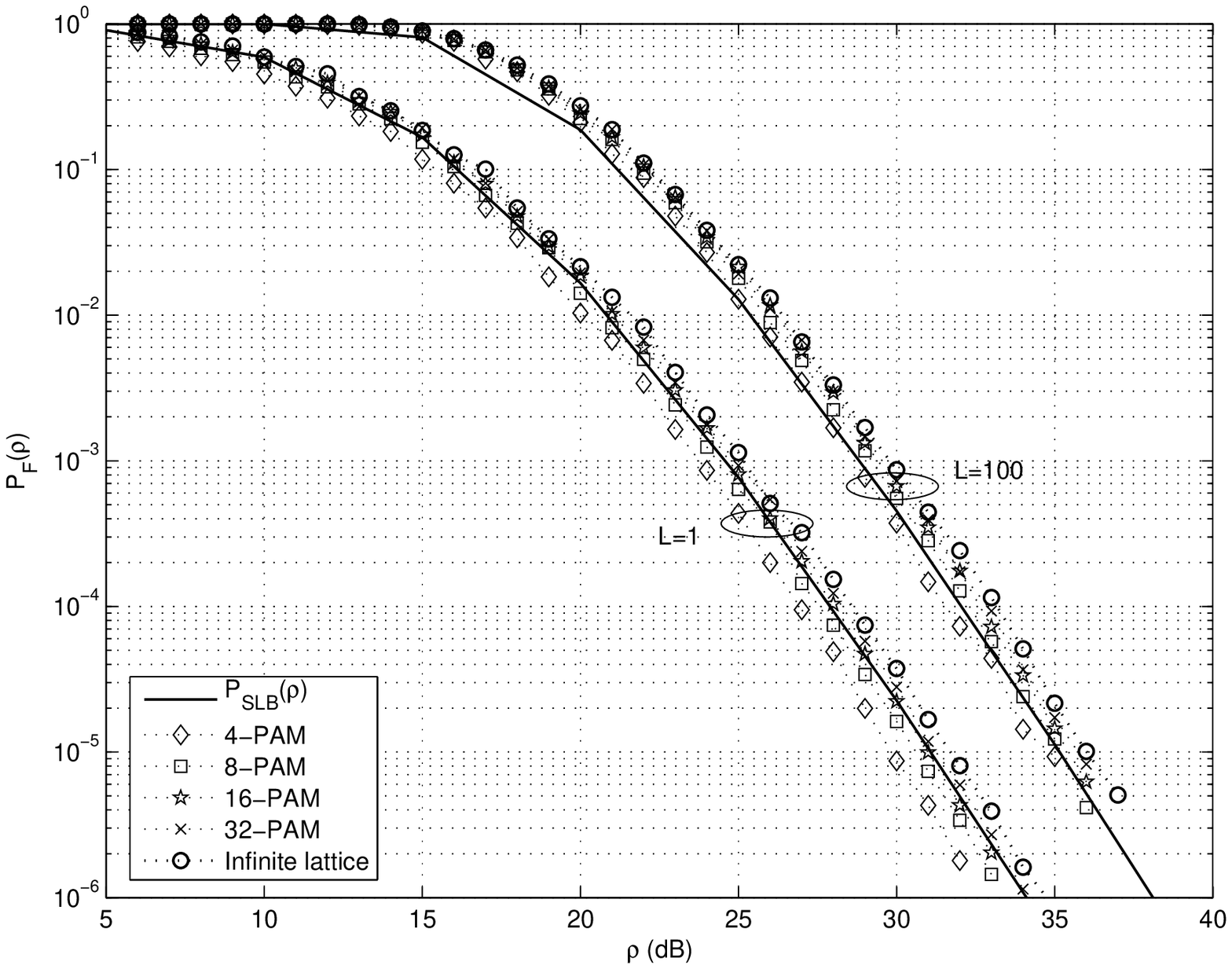}
  \caption{Frame error probability $P_{\rm f}(\rho)$ of the finite
  constellation generated with $4,8,16,32$-PAM, $P_{\rm f}(\rho)$
  with $N=4$ of the infinite lattice and sphere lower bound $P_{\rm slb}(\rho)$
  for $L=1,100$ with the Kr\"uskemper rotation.}
  \label{fig:peN4pam}
\end{figure}

\begin{figure}[htbp]
  \centering
  \includegraphics[width=0.9\columnwidth]{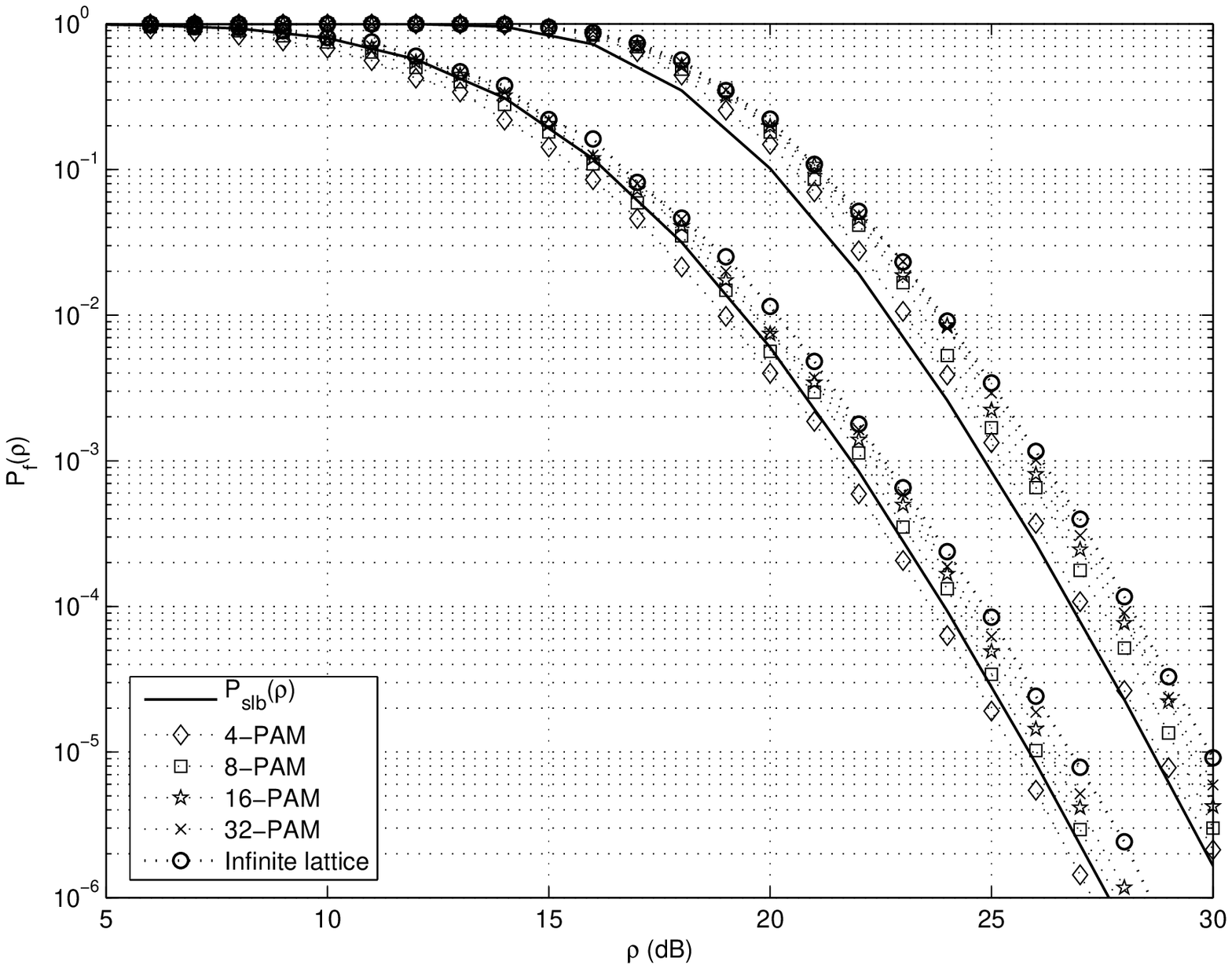}
  \caption{Frame error probability $P_{\rm f}(\rho)$ of the finite
  constellation generated with $4,8,16,32$-PAM, $P_{\rm f}(\rho)$
  with $N=8$ of the infinite lattice and sphere lower bound $P_{\rm slb}(\rho)$
  for $L=1,100$ with the cyclotomic rotation.}
  \label{fig:peN8pam}
\end{figure}

\begin{figure}[htbp]
  \centering
  \subfigure[\label{fig:pdfm1}$m=1$, $N=2,4,8,16,32,64$.]{\includegraphics[width=0.73\columnwidth]{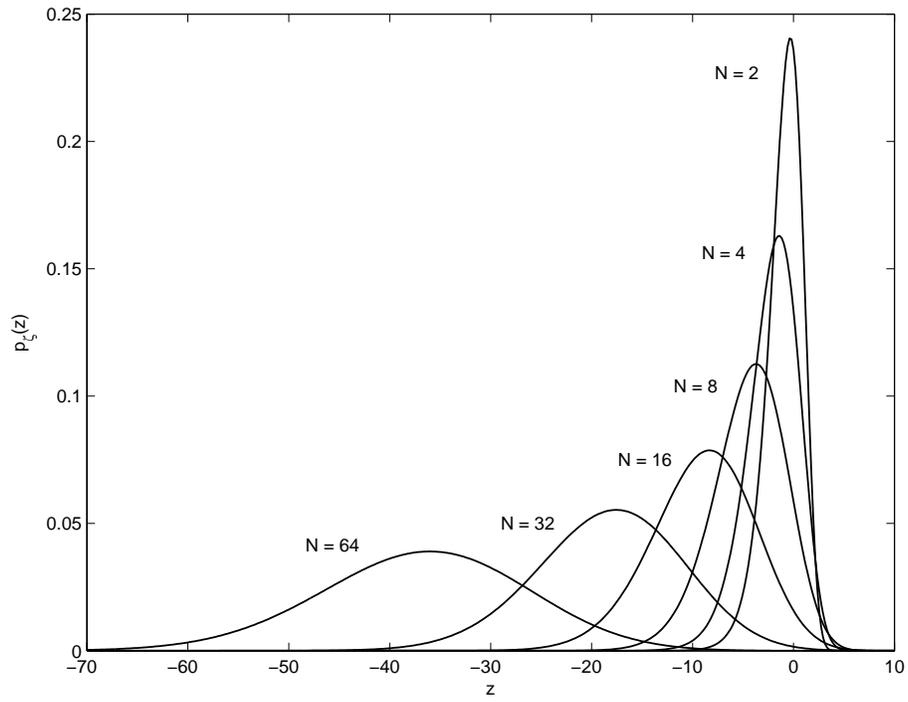}}
  \subfigure[\label{fig:pdfN8}$N=8$, $m=0.2,0.4,0.6,1,2$.]{\includegraphics[width=0.73\columnwidth]{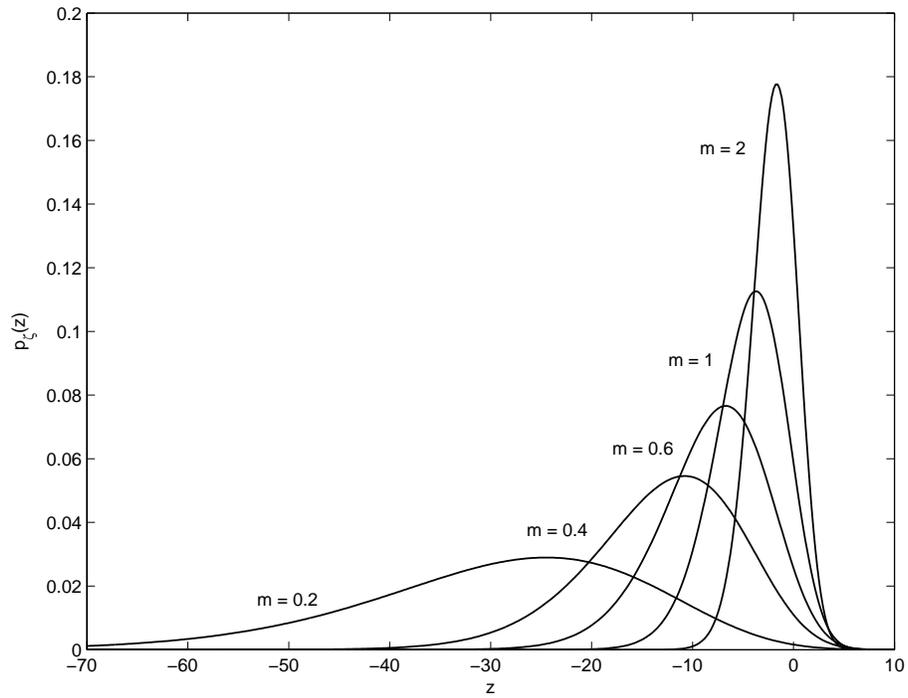}}
  \caption{Probability density function $p_\zeta(z)$ for various values of $N$ and $m$.}
    \label{fig:figpzN}
\end{figure}

\end{document}